# Noise-based deterministic logic and computing: a brief survey

Laszlo B. Kish [(1)], Sunil P. Khatri [(1)], Sergey M. Bezrukov [(2)], Ferdinand Peper [(3)], Zoltan Gingl [(4)], Tamas Horvath [(5,6)]

[(1)] *Department of Electrical and Computer Engineering, Texas A&M University, USA*

[(2)] *Laboratory of Physical and Structural Biology, Program in Physical Biology, NICHD, National Institutes of Health, Bethesda, MD 20892, USA*

[(3)] *National Institute of Information and Communications Technology, Kobe, 651-2492 Japan*

[(4)] *Department of Experimental Physics, University of Szeged, Dom ter 9, Szeged, H-6720, Hungary*

[(5)] *Department of Computer Science, University of Bonn, Germany*

[(6)] *Fraunhofer IAIS, Schloss Birlinghoven, D-53754 Sankt Augustin, Germany; email: tamas.horvath@iais.fraunhofer.de*

Abstract:

A short survey is provided about our recent explorations of the young topic of noise-based logic. After outlining the motivation behind noise-based computation schemes, we present a short summary of our ongoing efforts in the introduction, development and design of several noise-based deterministic multivalued logic schemes and elements. In particular, we describe classical, instantaneous, continuum, spike and random-telegraph-signal based schemes with applications such as circuits that emulate the brain's functioning and string verification via a slow communication channel.

## 1. Introduction: Errors (noise), speed (bandwidth) and power dissipation in computers

We present a short summary of our ongoing efforts in the introduction, development and design of several noise-based deterministic multivalued logic schemes and elements.

The numerous problems with current microprocessors and the struggle to continue to follow Moore's law [1-6] intensify the academic search for non-conventional computing alternatives. Among them, quantum computers have been proven to be disadvantageous for general-purpose computing [7,8] due to their error and energy dissipation problems; they are, at most, applicable for special-purpose-computing with a few specific applications. In general, quantum informatics is not the way to go for everyday use [8]. Today's computer logic circuitry is a system of coupled DC amplifier stages and this situation represents enhanced vulnerability against *variability* [1] of fabrication parameters such as threshold voltage inaccuracies in CMOS. Thin oxides imply great power dissipation due to leakage currents [1]. Thermal noise and its error generation, see Figure 1, and the related power dissipation limits are problems of fundamental nature [2-4]. If we want to increase the logic values in a CMOS gate while the error rate and clock frequency is kept constant, idealistically, with zero MOS threshold voltage, the necessary supply voltage scales with the number of logic values and the power dissipation scales with the square of the number of logic values, see Figure 1. Such a high price is unacceptable.

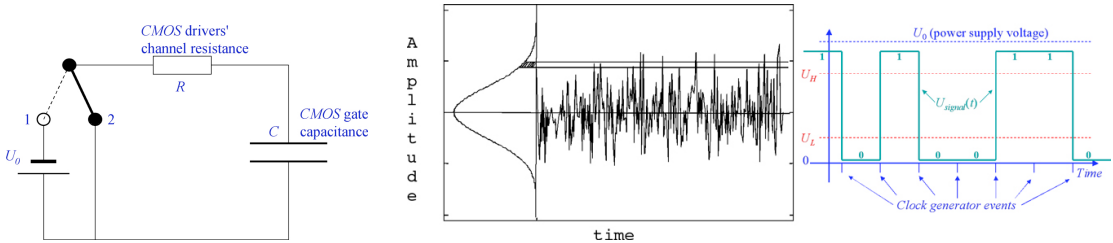

**Figure 1**. Left: Model for the MOSFET gate and its driving. Middle: Gaussian noise (thermal, 1/f, shot) and its amplitude distribution and its level-crossing events. Right: Critical levels in binary logic: when the signal is beyond $U_H$ it is interpreted as $H$ and when it is below $U_L$, it is interpreted as $L$.

Because the speed (bandwidth), energy dissipation and error probability are interrelated issues, it is important to emphasize that:

- Claims about *high performance* without *error rate* and *energy efficiency* aspects are interesting but meaningless for practical developments.

- Claims about *high energy efficiency* without *error rate* and *performance drop* aspects are interesting but meaningless for practical developments.

- Claims about *efficient error correction* without *energy requirement* and *performance drop* aspects are interesting but meaningless for practical developments.

- Finally, all these *performance-error-energy* implications *must be addressed at the system level* otherwise they are meaningless for practical developments. For example, *improvements at the single gate* level are interesting but unimportant if they are *lost at the system level*.

**2. Inspiration of noise-based logic: the brain**

Exploring *unconventional computing*, we can ask: How does biology do it??? Let us make a comparison between a digital computer and the brain in Table 1. We conjecture that the brain utilizes noise as information carrier. The result is a logic engine with a relatively low number of neurons (100 billions), extraordinary performance, acceptable error probability and extremely low power dissipation [9].

Motivated by these observations and the fact that *stochastic logic is very slow and produces too much errors*, our goal has been to study the possibility of constructing *deterministic high-performance logic schemes* utilizing noise as information carrier [10-14].

| Laptop computer | Human Brain |
|---|---|
| Processor dissipation: 40 W | Brain dissipation: 12-20 W |
| Deterministic digital signal | Stochastic signal: analog/digital? |
| Very high bandwidth (GHz range) | Low bandwidth (<100 Hz) |
| Sensitive for errors (freezing) | Error robust |
| Deterministic binary logic | Unknown logic |
| Potential-well based memory | Unknown memory mechanism |
| Addressed memory access | Associative memory access (?) |

**Table 1**. Comparison of digital computers and the brain

**3. Noise-based logic with continuum noise**

The first open question is what form of signal to utilize: a *continuum noise-based logic* [11,12], or a *logic with random spike trains* [13], like the brain? Some of the other concerns are listed here:

- Deterministic logic from noise: How to handle statistical averaging and the associated speed reduction?
- Speed in general: what is the potential speed gain due to the multivalued logic operations?
- Number of logic values: what is the most appropriate number of logic values in a single wire?
- Energy need: power dissipation versus performance?
- Error probability: how much errors can be tolerated and at what cost?
- Devices and logic gates: advantages and disadvantages of various realizations?

In Figure 2, the schematic of a continuum-noise-based logic gate is shown. The reference noises should be small to reduce power dissipation originating from their distribution. The scheme is very robust against the accumulation and propagation of fast switching errors. Thus the DC switches can be extremely poor quality with enormously high error probability, such as 0.1 . This situation helps to reduce switching power dissipation [11].

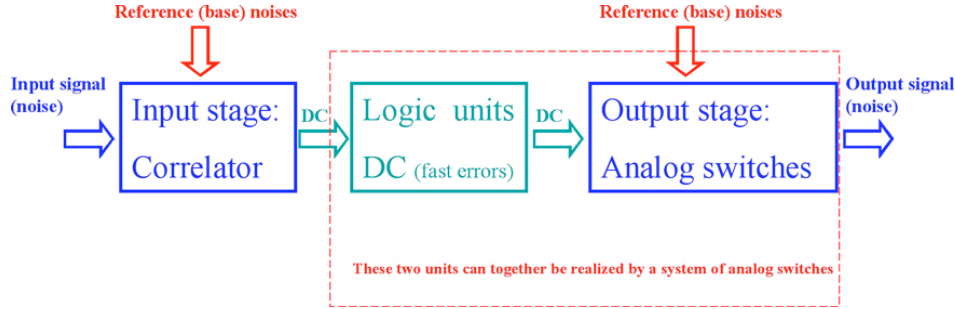

**Figure 2.** Schematics of continuum noise-based logic gates [11].

In Figure 3, the circuitry of the continuum-noise-based XOR logic gate is shown [11].

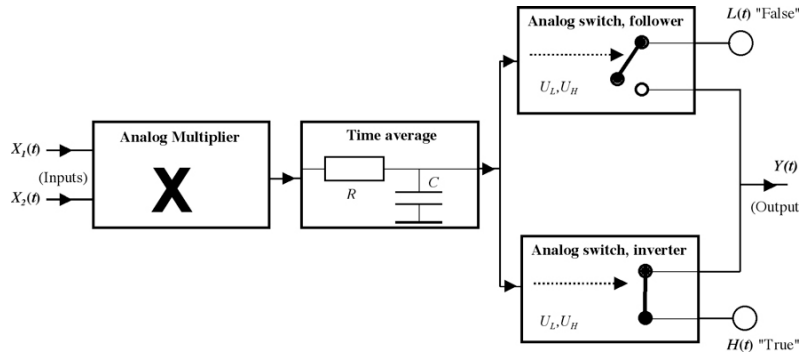

**Figure 3**. Circuitry of the continuum noise-based XOR logic gate.

The relevance of noise-based logic for nanoelectronics is obvious. In nanoelectronics the device size is smaller thus higher (small-signal) bandwidths are available. Also, more transistors are on the chip and significantly more noise.

Another key feature of continuum-noise-based logic is that a multidimensional logic hyperspace [11,12] can be built from the products of the original reference noises. We call the original reference noises *noise-bits* (similarly to the qubits of quantum computing) and the products *hyperspace* basis vectors. This hyperspace is equivalent to the Hilbert space of quantum informatics [12] however noise-based logic is classical physical therefore it operates without the limitations of quantum-collapse of wavefunctions. Figure 4 illustrates how $2^N$ orthogonal bit vectors can be created by using $2N$ noise-bits (left) and how $2^{N-1}$ orthogonal bit vectors can be created by using $N$ noise-bits. In [12] a noise-based string search algorithm with faster speed than Grover's quantum search algorithm is shown. With practical data, the noise-based engine has the same hardware complexity class as the quantum engine.

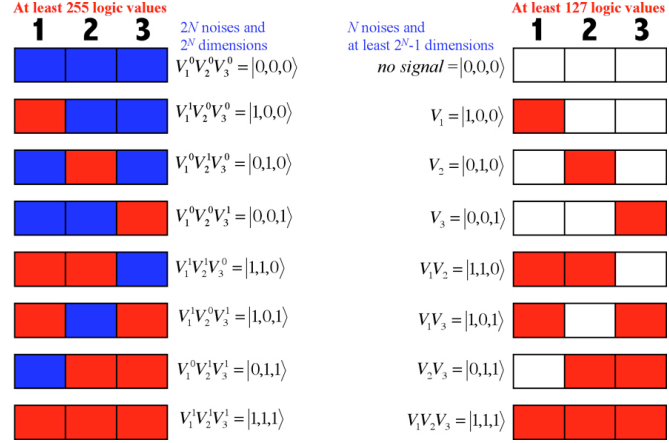

**Figure 4**. Multidimensional logic hyperspace can be built from the products of the original reference noises [11,12]. Left: complete quantum Hilbert space emulation including vacuum states [12] with three noise-bits. The zero elements have their own noise. Right: hyperspace emulation with three simplified noise-bits where the zero element represented by nonexistent corresponding noise term in the product [12].

## 4. Some of the advantages and disadvantages of continuum noise-based logic

The continuum-noise-based logic has the following potential advantages [11,12]:

*i*. Arbitrary number of logic values can be transmitted in a single wire. Utilization of quantum-like hyperspace is possible.

*ii*. Due to the zero mean of the stochastic processes, the logic values are AC signals and AC coupling can ensure that variability-related vulnerabilities are strongly reduced.

*iii*. The scheme is robust against noises and interference. The different basic logic values are orthogonal not only to each other but also to any transients/spikes or any background noise including thermal noise or circuit noise, such as 1/f, shot, generation-recombination, etc., processes. Moreover, the usual binary switching errors do not propagate and accumulate.

*iv*. Due to the orthogonality and AC aspects (points ii and iii), the logic signal on the data bus can have very small effective amplitude (approaching the background noise). This property potentially allows a significant reduction of supply voltage. This characteristics and the robustness against switching errors, have a potential to reduce the energy consumption.

Disadvantages:

*a*. The (continuum noise-based) noise-based logic is slower due to the need of averaging at the output of the correlator, see Figure 2. In binary gates, its digital bandwidth is about 1/500 of the bandwidth of the small-signal-bandwidth of the noise [11]. That yields a clock frequency of about 0.1-0.5 GHz with today's chip technology. This deficiency may be compensated by the multivalued logic abilities which are not available for traditional binary gates.

*b*. The need for more complex hardware. This deficiency may also be compensated by the multivalued logic abilities.

*c*. Hardware simulations [14] with sinusoidal signals instead of noise indicate that CMOS technology may not be the best implementation of continuum noise-based logic because of the low transconductance, the non-zero threshold voltage and the large crossbar currents resulting in extra power dissipation. Exponential devices such as single electron transistors or bipolar devices may be more appropriate. Again, this deficiency seen with the tested CMOS architectures [14] may also be compensated by the multivalued logic abilities.

Finally, a quick comparison with quantum computing [12]:

- The approach can implement binary or multivalued logic, with optional superposition of states (like quantum logic states).

- Entanglement can also be made in the superposition (similarly to quantum states).

- However, in the noise-based scheme, collapse of the wavefuntion does not exist.

- All the superposition components are accessible at all times (much better situation than general-purpose multivalued quantum logic applications where the calculations must be repeated many times to generate and extract statistics )

- A noise-based string search algorithm outperforming Grovers quantum search is proposed with the same hardware complexity as the quantum engine when it is used for real data [12].

**5. Noise-based logic with random spikes [13]**

In this section, we discuss the deterministic logic scheme with random spike trains [13] which is conjectured to be a simple model of brain logic. Neural signals are stochastic unipolar spike trains: their product has non-zero mean value. This situation rules out multiplication as a means to building a hyperspace. Thus the basic question is: Can we build an orthogonal noise-bit type multidimensional Hilbert space using an alternate approach? The answer is yes, by using set-theoretical operations between partially overlapping neural spike trains which we call *neuro-bits*. For two and three neurobit examples, see Figure 5.

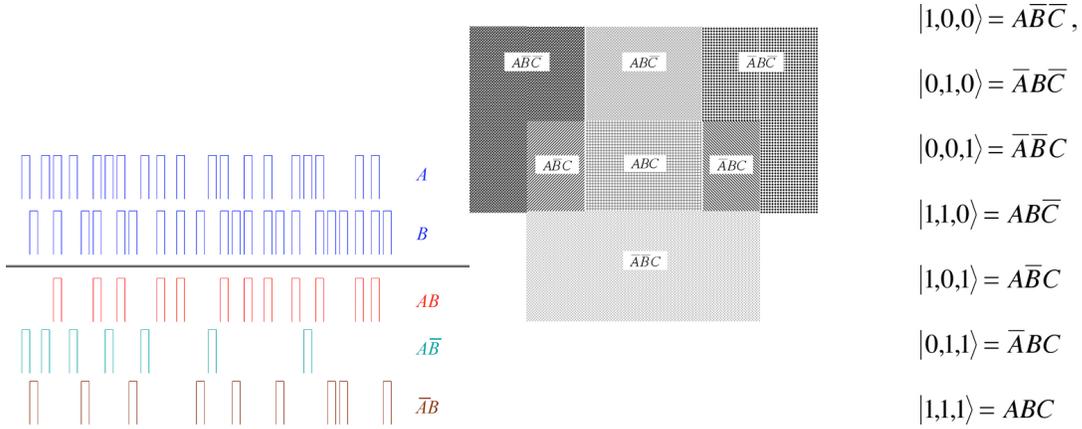

$$|1,0,0\rangle = A\overline{B}\overline{C},$$
$$|0,1,0\rangle = \overline{A}B\overline{C}$$
$$|0,0,1\rangle = \overline{A}\overline{B}C$$
$$|1,1,0\rangle = AB\overline{C}$$
$$|1,0,1\rangle = A\overline{B}C$$
$$|0,1,1\rangle = \overline{A}BC$$
$$|1,1,1\rangle = ABC$$

**Figure 5**. Left: three-dimensional hyperscape generated from two neuro-bits (partially overlapping neural spike trains) utilizing set-theoretical operations [13]. Right and middle: the same with three neuro-bits and seven-dimensional hyperscape, and its illustration by overlapping squares as neuro-bits.

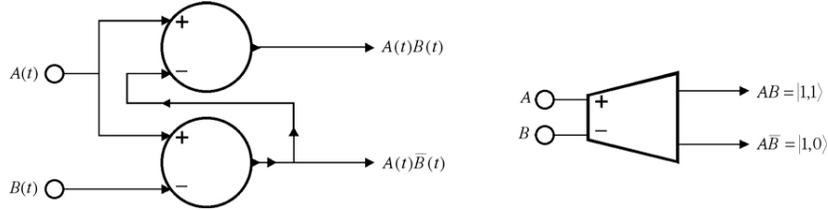

**Figure 6**. The orthon [13] consists of two neurons and has two inputs and two outputs. Left: its neural circuitry where the plus stands for the excitatory input and the minus for the inhibitory input of the neuron. Right: Its symbol.

To address the issue of realizing such a hyperspace using neurons, we present a key element, the *orthon* [13] which consists of two neurons (see Figure 6). The *orthogonator* unit, which prepares the multidimensional hyperspace from the neuro-bits (see Figure 7) can be built of orthons and neurons. If an orthogonator has $N$ inputs, it has $2^N$-1 orthogonal outputs. There are two families of orthogonators [13]:

*i. Nonsequential orthogonator* (for neural systems): Such a circuitry [13] has $N$ parallel inputs run by parallel, partially overlapping random spikes trains and it generates the available set-theoretical intersections of the input spikes. There are $M = 2^N - 1$ output wires with non-overlapping spike trains [13]. The advantage of the nonsequential orthogonator is that it may be faster and it can transform a set of partially overlapping spike trains into a set of orthogonal trains. Figure 7 illustrates a non-sequential orthogonator which produces $2^3-1=7$ non-overlapping spike trains at its outputs from 3 neurobits driving its inputs.

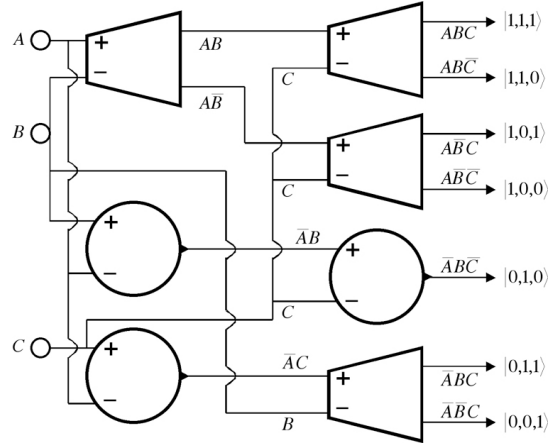

**Figure 7**. Neural orthogonator [13]. Third order, non-sequential.

ii) *Sequential orthogonator* for chip applications [15]: It has a single input to be fed by a single (infinite) spike sequence. A *controlled rotary switch* will distribute the subsequent spikes to *M* output wires in a cyclic way (by demultiplexing). This circuitry may be more advantageous in certain chip designs [15] however it lacks the random fingerprint like nature of the output spike trains and the related resilience.

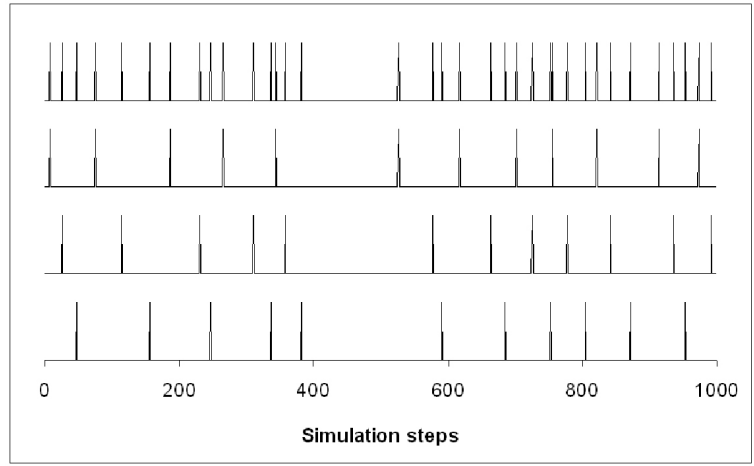

**Figure 8**. Input and output signals of second-order sequential orthogonator [15] driven by zero-crossing events of band-limited white noise. Upper line: the original spike train. Lower lines: the orthogonal sub-trains at the three outputs.

The great potential of this logic scheme comes from the fact that, when analyzing multidimensional superpositions, the very first spike of a searched hyperspace base vector will indicate that this vector is an element of the superposition, see Figure 9. The lack of this spike will similarly be informative. We call such an operation neural Fourier transformation [13], see its circuitry in Figure 9. Even though the logic signal is noise, no time averaging is needed, just some delay until the first relevant spike appears (or is found to be missing).

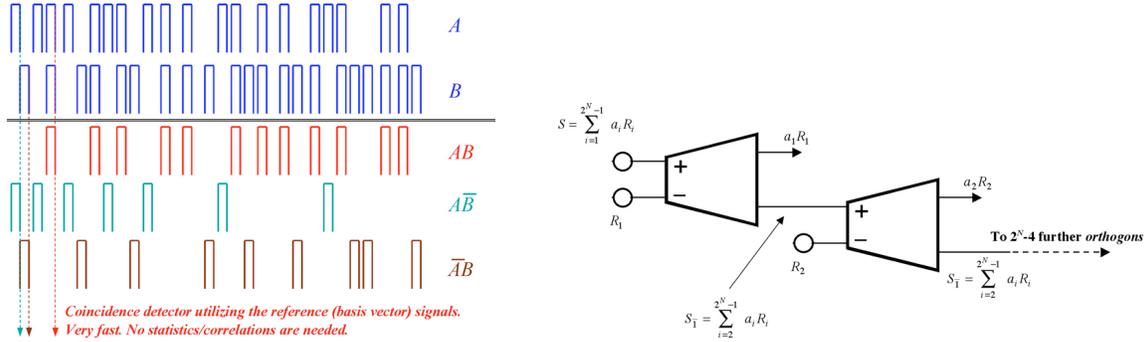

**Figure 9**. Left: the principle of neural Fourier transformation. The firstly arriving spike of that vector will indicate the presence of that vector in the superposition. Left: detecting the orthogonal elements of a two neuro-bits system. Right: the circuitry of neural Fourier transformation [13].

The essential open question is about this scheme is whether these circuit elements are found in the brain? There are several encouraging findings in this respect. Observations of the role of the timing of single spikes in sensory neural signals (called sometimes first-spike-coding) [16,17] and the recently observed decorrelated spike trains in the activity of cortical neurons [18], suggest that the brain may be using similar logic .

**6. Instantaneous noise-based logic [19]**

Instantaneous noise-based logics are Boolean logic schemes that do not require a time-averaging process. We prove that these logics are universal by showing that they can realize AND and NOT gates. These logic systems have the potential to do fast noise-based logic operations with low power consumption in special-purpose engines. However, to interface them with other types of logic, a classical noise-based logic (with its time average units) is needed. Thus interfacing the final results will be slower but the computation process is extremely fast [19].

*6.1 Instantaneous logic by random telegraph waves*

Let us suppose that the noise function $H(t)$ representing the logic value $H$ is a random telegraph wave (RTW) defined as follows. The absolute value of the amplitude is 1, and at the beginning of each time step, the amplitude changes its sign with probability 1/2. Thus the RTW wave is a square wave with 50% probability to have a +1 value, and the same probability to have -1 amplitude.

The NOT gate operation is defined by Equation 1, and its hardware is a linear differential amplifier shown in Figure 10.

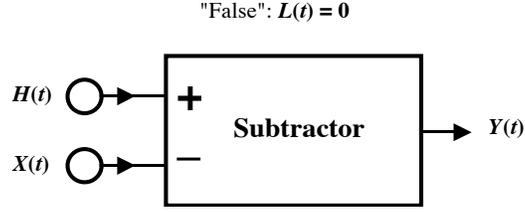

**Figure 10**. The NOT (INVERTER) gate [19]. It consists of a linear amplifier.

$$Y(t) = \text{NOT } X(t) = H(t) - X(t) \tag{1}$$

If $X(t) = L(t) = 0$ then $Y(t) = H(t)$. If $X(t) = H(t)$, then $Y(t) = 0 = L(t)$.

The AND gate operation is defined by Equation 2 and it's hardware consists of two multipliers as shown in Figure 11:

$$Y(t) = X_1(t) \text{ AND } X_2(t) = X_1(t)X_2(t)H(t) \tag{2}$$

The output will produce $H(t)$ when both inputs have $H(t)$ signal. Otherwise the output produces $L(t)$. Note that the multipliers can be simplified to linear amplifiers (with amplification of +1 and -1, respectively) and switches because of the trivial tasks of multiplication by ±1.

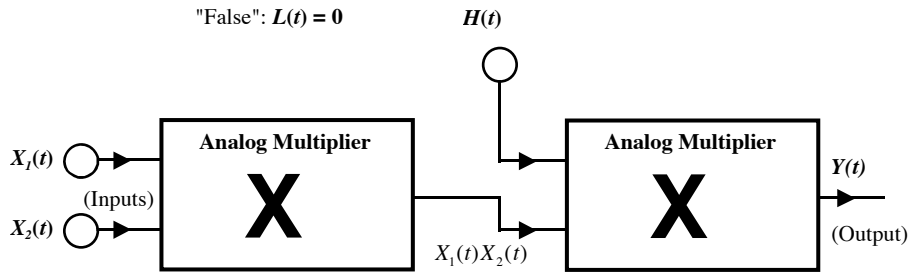

**Figure 11**. The AND gate [19].

Since both the AND and NOT gates can be realized, the RTW-based *INBL* scheme defined above is universal.

## 6.2 Instantaneous logic based on random spike sequences

This logic [19] is the *binary version* of the brain logic scheme described in Section 5 and originally introduced in [13] as a hyperspace. The spike train representing the logic value $H$ is the spike train $H(t) \neq 0$ and the logic value $L$ is represented by no spikes

($L(t) = 0$). To prove universality, we show the NOT and the AND gates below. The NOT gate is defined as:

$$Y(t) = \text{NOT } X(t) = H(t) \cap \overline{X}(t) \qquad (3)$$

For $X(t) = H(t)$, $Y(t) = L(t) = 0$ and for $X(t) = L(t) = 0$, $Y(t) = H(t)$.

The AND gate is defined as:

$$Y(t) = X_1(t) \text{ AND } X_2(t) = X_1(t) \cap X_2(t) \qquad (4)$$

The orthon [13] based representation of the NOT and AND gates are shown in Figure 4. The orthon can be used to perform both logic functions.

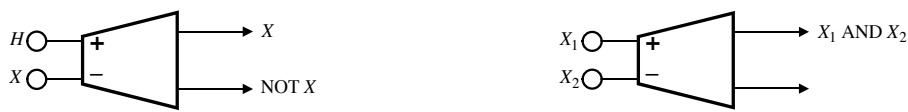

**Figure 12**. The binary NOT (left) and AND (right) gates utilize the orthon element of spike-based logic [19]. Note that the upper output is not used in the circuit on the left, and the lower output is not used in the circuit on the right.

## 7. Efficient string verification by instantaneous noise-based logic [20]

In [20], we showed two string verification methods with low communication complexity, using the noise-based hyperspace. The idea was to verify if two *N*-bit strings are identical. One string is held by Alice and the other by Bob. Alice and Bob would like to test if their strings are different, with minimum communication complexity. One scheme was based on continuum noise-based logic and the other on RTW-based logic. The idea is to represent each possible bit value in the *N*-bit string with 2*N* independent reference noise processes that are identical for Alice and Bob. Then the signal to be communicated is the product of the noises representing the actual bit values of the *N*-bit string; it is a hyperspace vector containing information about all the bits. If even one of the bit values is different at Alice and Bob, the resulting signals at the two sides are independent stochastic processes. By communicating, for example, only 83 signal bits, Alice and Bob can determine with $2^{-83} \approx 10^{-25}$ error probability that the two *N*-bit strings with arbitrary length are different. Notice that the error probability of an idealistic gate in today's computer is similar to this value [2] which means that communicating more bits for this purpose is meaningless. This RTW-based operation could also be interpreted as calculating hash functions by noise-based logic.

## 8. Final remarks

When discussing the operation of the brain, we must realize that even though the human brain is extremely powerful compared to a computer, there are areas where computers are more efficient: tasks where a large number of primitive operations must be done quickly with high accuracy are better performed by a computer. The brain is powerful where computers are weak: quick identification of complex situations and patterns or incomplete patterns (part of intuition and intelligent decision). These are done with certain level of error tolerance. Thus, it is possible that the real applications of noise-based logic will be hybrid engines in future intelligent computers and robots, where the classical computing part will stay as a supporting framework.

**Acknowledgements**

Discussions with Kalyan Bollapalli and Swaminathan Sethuraman are appreciated. Work by Z.G. was supported by grant OTKA K69018 of the Hungarian Academy of Sciences.

**References**

[1] S. Khatri and N. Shenoy, "Logic Synthesis", in "EDA for IC Implementation, Circuit Design and Process Technology", Editors: L. Lavagno, L. Sheffer, G. Martin, CRC Press (2006), ISBN 978-0849379246.
[2] L.B. Kish, "End of Moore's Law; Thermal (Noise) Death of Integration in Micro and Nano Electronics", *Physics Letters A* **305** (2002) 144–149.
[3] W. Porod, R.O. Grondin, D.K. Ferry, Dissipation in computation, *Physical Review Letters* **52** (1984) 232-235.
[4] R.K. Cavin, V.V. Zhirnov and J.A. Hutcby, "Energy barriers, demons and minimum energy operation of electronic devices", *Fluctuation and Noise Letters* **5** (2005) C29-C38.
[5] M. Forshaw, R. Stadler, D. Crawley and K. Nikoli, "A short review of nanoelectronic architectures", *Nanotechnology* **15** (2004) 220–223.
[6] P. Korkmaz, B.E.S. Akgul, K.V. Palem, "Energy, performance, and probability tradeoffs for energy-efficient probabilistic CMOS circuits", IEEE Transactions on circuits and systems **55** (2008) 2249-2262.
[7] J. Gea-Banacloche, L.B. Kish, "Comparison of Energy requirements for Classical and Quantum Information Processing", *Fluctuation and Noise Letters* **3** (2003) C3-C6.
[8] J. Gea-Banacloche and L.B. Kish, "Future directions in electronic computing and information processing", *Proceedings of the IEEE* **93** (2005) 1858-1863.
[9] S.M. Bezrukov and L.B. Kish, "How much power does neural signal propagation need?", *Smart Mater. Struct.* **11** (2002) 800–803.
[10] L.B. Kish, "Thermal noise driven computing", *Applied Physics Letters* **89** (2006) 144104; http://arxiv.org/abs/physics/0607007
[11] L.B. Kish, "Noise-based logic: binary, multi-valued, or fuzzy, with optional

superposition of logic states", *Physics Letters A* **373** (2009) 911-918; http://arxiv.org/abs/0808.3162

[12] L.B. Kish, S. Khatri, S. Sethuraman, "Noise-based logic hyperspace with the superposition of $2^N$ states in a single wire", *Physics Letters A* **373** (2009) 1928-1934; http://arxiv.org/abs/0901.3947

[13] S.M. Bezrukov, L.B. Kish, "Deterministic multivalued logic scheme for information processing and routing in the brain", *Physics Letters A* **373** (2009) 2338-2342, http://arxiv.org/abs/0902.2033 .

[14] K. Bollapalli, S. Khatri, L.B. Kish, "Implementing Digital Logic with Sinusoidal Supplies", Design Automation and Test in Europe conference, Dresden, Germany, March 8-12, 2010.

[15] Z. Gingl, S. Khatri, L.B. Kish, "Toward brain-inspired computing", submitted; http://arxiv.org/abs/1003.3932 .

[16] R.S. Johansson, I. Birznieks, "First spikes in ensembles of human tactile afferents code complex spatial fingertip events", *Nature Neurosci* **7** (2004) 170-7.

[17] R. Guyonneau, R. VanRullen, S.J. Thorpe, "Neurons tune to the earliest spikes through STDP", *Neural Computation* **17** (2005) 859-879.

[18] A.S. Ecker, P. Berens, G.A. Keliris, M. Bethge, N.K. Logothetis, A.S. Tolias, "Decorrelated neuronal firing in cortical microcircuits", *Science* **327** (2010) 584-587.

[19] L.B. Kish, S. Khatri, F. Peper, "Instantaneous noise-based logic", *Fluctuation and Noise Letters* **9** (December 2010); http://arxiv.org/abs/1004.2652 .

[20] L.B. Kish, S. Khatri, T. Horvath, "Computation using Noise-based Logic: Efficient String Verification over a Slow Communication Channel", submitted; http://arxiv.org/abs/1005.1560 .